\documentclass[compactaffiliation]{Interspeech}

\usepackage{xcolor}
\newcommand{\myRed}[1]{\textcolor{black}{#1}}

\interspeechcameraready

\title{Novel Loss-Enhanced Universal Adversarial Patches for Sustainable Speaker Privacy}

\author[affiliation={1,2,3}]{Elvir}{Karimov}
\author[affiliation={5,2,3}]{Alexander}{Varlamov}
\author[affiliation={3,4}]{Danil}{Ivanov}
\author[affiliation={1,3,2}]{Dmitrii}{Korzh}
\author[affiliation={1,2,3}]{Oleg Y.}{Rogov}

\affiliation{RSI Group}{AIRI}{Moscow, Russia}
\affiliation{Safe AI Lab}{MTUCI}{Moscow, Russia}
\affiliation{}{Skoltech}{Moscow, Russia}
\affiliation{}{Vein CV LLC}{Moscow, Russia}
\affiliation{Kandinsky Lab}{Sber}{Moscow, Russia}
\email{karimov@airi.net, korzh@airi.net, rogov@airi.net}

\usepackage{comment}
\usepackage{tikz}

\usepackage{amsmath}

\usepackage{latexsym}
\usepackage{amssymb}
\usepackage{amsthm}
\usepackage{booktabs}
\usepackage{enumitem}
\usepackage{microtype}
\usepackage{tabularx}
\usepackage{mathtools}
\usepackage{subcaption}
\usepackage[noend]{algpseudocode}

\theoremstyle{plain}
\newtheorem{theorem}{Theorem}

\theoremstyle{definition}

\theoremstyle{remark}

\DeclareMathOperator*{\argmin}{argmin}
\DeclareMathOperator*{\argmax}{argmax}
\usepackage{xcolor}
\usepackage{algorithm}

\begin{document}

\maketitle

\begin{abstract}
Deep learning voice models are commonly used nowadays, but the safety processing of personal data, such as human identity and speech content, remains suspicious. To prevent malicious user identification, speaker anonymization methods were proposed. Current methods, particularly based on universal adversarial patch (UAP) applications, have drawbacks such as significant degradation of audio quality, decreased speech recognition quality, low transferability across different voice biometrics models, and performance dependence on the input audio length. To mitigate these drawbacks, in this work, we introduce and leverage the novel Exponential Total Variance (TV) loss function and provide experimental evidence that it positively affects UAP strength and imperceptibility. Moreover, we present a novel scalable UAP insertion procedure and demonstrate its uniformly high performance for various audio lengths. 
\end{abstract}

\section{Introduction}
\label{s:intro}
Speaker recognition (SR) systems: voice-biometrics, automatic speaker identification (ASI) \cite{desplanques2020ecapa, wang2023cam++} have become increasingly prevalent in modern voice-driven applications, from intelligent assistants to biometric security. These networks transform each speech utterance into a fixed-dimensional embedding vector, which encodes the unique characteristics of the speaker's identity.

However, the growing use of voice data raises significant privacy concerns, particularly the risk of biometric identity theft, voice cloning, and unauthorized speaker recognition, which could compromise personal security. While these systems offer convenience and security, they expose users to the risk of being tracked without their consent. As a result, protecting speaker privacy has emerged as a critical issue, leading to the exploration of methods such as speaker anonymization and adversarial attacks to mislead malicious biometric systems.

Speaker anonymization systems (SAS) \cite{deng2023v, xie2021enabling, zhang2023imperceptible} are designed to \myRed{confuse} ASI models while maintaining the intelligibility, content, and naturalness of \myRed{modified} audio. In addition, they ensure that downstream tasks, such as speech recognition and classification, remain unaffected.

In this work, we focus on the improvement of SAS based on additive adversarial attacks \cite{szegedy2013intriguing,Korzh2025} and, particularly, on universal adversarial patches (UAP) \cite{moosavi2017universal, xie2021enabling, israel}. 

Let $f$ be a speaker recognition model that maps an input audio to a speaker embedding space $f: \mathbb{R}^n \to \mathbb{R}^d$. Given an audio sample $x \in \mathbb{R}^n$ and the set of  enrolled speakers representations $e_i \in \mathbb{R}^d$, the model assigns the speaker identity using cosine similarity function $\rho$:  

\begin{equation}
i = \argmax_k \rho(f(x), e_k).    
\end{equation}

We introduce a universal adversarial perturbation (UAP), $\hat \delta \in \mathbb{R}^{l}$, that aims to mislead the speaker recognition system for any speaker:
\begin{equation}
\argmax_k \rho(f(x + \delta), e_k) \neq \argmax_k \rho(f(x), e_k),
\end{equation}
where $\delta \in \mathbb{R}^n$ is a repeated UAP to suit the length of $x$.

\textbf{Our contributions can be summarized as follows:}
\begin{itemize}
    \item \textbf{Incorporation of the novel loss function.} We propose a novel Exponential TV loss function inspired by TV loss from the image domain, designed to preserve the imperceptibility of UAPs.
    \item \textbf{Length-Independent UAP.} We introduce a length-independent UAP generation approach by training on long audio samples with a repeat padding strategy, making it effective for real-world applications. \myRed{To the best of our knowledge, this strategy, although being well-known, was not used in prior UAP training.}
    \item \textbf{\myRed{Length-Agnostic} Evaluation Procedure.} We establish a rigorous evaluation protocol that accounts for dataset biases, including variations in loudness levels. Furthermore, a \myRed{proper} padding strategy based on audio repetition is implemented to prevent the UAP from exploiting artificially silent segments, ensuring robustness across different audio lengths.
\end{itemize}

\subsection{Related Work}

Recently, the accuracy and efficiency of automatic speaker identification and verification systems (ASI, ASV) \cite{ snyder2018x, jung2019RawNet, Borodin2024} have been drastically enhanced. A notable breakthrough in this field is the development of the x-vector system \cite{snyder2018x}, built upon Time Delay Neural Network (TDNN) architecture. The x-vector framework leverages one-dimensional convolutional layers to capture temporal dependencies in speech signals effectively. This approach was improved in the ECAPA-TDNN model \cite{desplanques2020ecapa} that extends the TDNN by employing squeeze-and-excitation blocks and multi-scale features, allowing the model to capture long-range temporal dependencies better. ECAPA-TDNN integrates features from multiple time scales by iteratively aggregating information from previous layers, thus refining the representation of speech segments for more accurate speaker embeddings. Further advancements include the development of the D-TDNN (Densely Connected TDNN), Context-Aware Masking (CAM) module, resulting in CAM++ \cite{wang2023cam++, wang2023wespeaker} model which significantly improves key verification metrics and inference time. 

SAS \cite{liu2024transferable, NeekharaHPDMK19,  ZhangDTWG23, LuoSLX21, waveunet} aims to conceal identity properties while preserving the extraction of other information from the speech utterance. There are several ways to accomplish this. The most straightforward way is to transcribe speech with the automatic speech recognition (ASR, STT) \cite{gulati2020conformer, radford2023robust} model and then utter the text via the text-to-speech \cite{shen2018natural} model with a dummy voice. However, this approach distorts voice style features, has a significant computational overhead, and is barely applicable in real-time streaming applications. The development of this approach involves extracting the speaker's internal voice features and swapping them with random or mean features of several speakers, as well as audio regeneration, which can preserve the person's style of speech but hide the identity. 

Audio-specific signal filters \cite{miao2023speaker, miao2022analyzing} might be applied to \myRed{confuse} ASI, distorting specific audio features. However, such approaches drastically degrade the audio quality. In contrast, one can use generative models to imperceptibly adjust the audio in a non-additive manner. For instance, V-Cloak \cite{deng2023v} iteratively creates a unique set of adversarial perturbations for a particular input via WaveUnet \cite{waveunet}. VSMask \cite{wang2023vsmask} accelerates the creation of adversarial perturbations for the next chunk of audio in streaming. Another prominent way is the creation of additive adversarial perturbation, which can be considered an adversarial attack \cite{szegedy2013intriguing, goodfellow2014explaining, madry2017towards}, that can be insignificant to the human's ear or eye but can drastically affect the deep learning model's performance. Additionally, one can apply UAPs \cite{xie2021enabling, zhang2023imperceptible, liu2024transferable, israel}, which are suitable for fooling the recognition of different audio of different speakers and sometimes even for the different ASI models simultaneously. In an over-the-air scenario, if appropriately trained, the UAP can be emitted continuously to ensure ongoing real-world and real-time anonymization.

\section{Methodology}
Our work focuses on developing a training procedure and a \myRed{length-agnostic} testing approach for constructing robust UAPs that should fail ASI for unseen speakers and various audio lengths.

\subsection{Preliminaries}
In a typical speaker recognition system, one uses a neural network
$f: \mathbb{R}^{L} \rightarrow\mathbb{R}^{d}$, where $L$ is an upper bound on the length of the input audio waveform (after padding or truncation), and $d$ is the embedding dimension. For any audio input $x \in \mathcal{X} \subset \mathbb{R}^{L}$,
the network $f$ outputs a vector $f(x) \in \mathbb{R}^{d}$,
called the \emph{speaker embedding}. It can be utilized for speaker identification or verification tasks by comparing embeddings using a suitable distance measure, e.g., Euclidean or cosine distance, or it can further be processed for classification. 
We denote $\mathcal{B}$ as decision boundary of $f$, i.e., $\{z \in \mathbb{R}^L : f(z) \text{ changes class}\}$. The perturbed input on $\mathcal{B}$ is then $z = x + r(x)$, where $x$ is the original input, and  the classification margin at $z$ is  $\mathcal{M}(z) = \mathrm{distance}(z, \mathcal{B})$ and $r(x) = \text{argmin}_{r} ||r||_2$ s.t. $f(x+r) \neq f(x)$ is the minimal adversarial perturbation.

We denote a shorter patch length as $l \leq L $ and therefore, the UAP $\hat \delta \in \mathbb{R}^{l}$ is a time-domain perturbation of length $l$ intended to be \textit{repeated} or \textit{tiled} across any audio signal of length $n \leq L$.
 
If $l$ divides $n$, the final length is precisely $n$. The resulting additive perturbation can be interpreted as broadcasting $\hat \delta$ continuously across the time axis of the audio. 
For the UAP construction we introduce $\Theta(\hat{\delta}) = (\hat{\delta_1}, \dots, \hat{\delta_l}, \dots)_{1:L}$ which is the tiling operator repeating $\hat{\delta} \in \mathbb{R}^l$.  

We define the maximum perturbation $\epsilon > 0$:  norm $||\hat{\delta}|| \leq \epsilon$ and $\beta \in (0,1)$ as the probability tolerance for a universal attack.

Boundary  $\mathcal{B}$ of a classifier
$\Phi : \mathbb{R}^{d} \to \{1,\dots,K\}$ in $\mathbb{R}^{d}$ for speaker recognition models
can exhibit \emph{positive curvature} \cite{moosavi2017universal,Rahmati2020GeoDAAG}, meaning small steps in certain directions cause boundary crossings (misclassification). Universal perturbations arise when multiple inputs share such high-curvature directions. Tiling a short patch $\hat \delta$ across the time axis leverages these directions to degrade or \myRed{mislead} the recognition system for many utterances.

\begin{theorem}[UAP Generalization]
\label{thm:UAP}
Suppose there exists a subspace $S \subset \mathbb{R}^d$ ($\dim S = m \ll d$) such that for most $x \in \mathcal{X}$ the Hessian $H_z = \nabla^2 \mathcal{M}(z)$ of the margin at $z = x + r(x)$ satisfies:
\begin{equation}
\inf_{\substack{\mathbf{v} \in S \\ ||\mathbf{v}||=1}} \mathbf{v}^\top H_z \mathbf{v} \geq \kappa.
\end{equation}
And $J_f(x)$ maps tiled perturbations to $S$:
    \begin{equation}
    || \mathrm{Proj}_S(J_f(x)\cdot \Theta(\hat{\delta})) || \geq c || J_f(x) \cdot \Theta(\hat{\delta}) || \quad (c > 0).
    \end{equation}

Then $\forall \beta \in (0,1)$, $\exists\hat{\delta} \in \mathbb{R}^l$ with $||\hat{\delta}|| \leq \epsilon$ such that:
\begin{equation}
\mathbb{P}_{x \sim \mathcal{X}}\big(f(x + \Theta(\hat{\delta})) \neq f(x)\big) \geq 1 - \beta - e^{\left(-\frac{m \kappa^2 \epsilon^2}{2 \sigma^2}\right)},
\end{equation}
where $\sigma^2$ bounds the variance of $J_f(x)\cdot \Theta(\hat{\delta})$ over $x$, $\kappa > 0$ is the minimum curvature threshold in $S$.
\end{theorem}

\begin{proof}
We begin by noting that $r(x)$ and at $z = x + r(x)$, the unit normal $\mathbf{n}(x)$ to $\mathcal{B}$ satisfies $\langle r(x), \mathbf{n}(x) \rangle = ||r(x)||_2$.  
\begin{equation}
r(x) = \argmin_{r \in \mathbb{R}^L} ||r||_2 \quad \text{s.t.} \quad f(x + r) \neq f(x).
\end{equation} 
The margin function $\mathcal{M}(z) = \mathrm{distance}(z, \mathcal{B})$ has Hessian $H_z = \nabla^2 \mathcal{M}(z)$, encoding boundary curvature.  For $\mathbf{v} \in S$, the second-order Taylor expansion of $\mathcal{M}$ gives:  
\begin{equation}
\label{eq:expansion}
\mathcal{M}(z + \mathbf{v}) \approx \mathcal{M}(z) + \underbrace{\nabla \mathcal{M}(z)^\top \mathbf{v}}_{=0 \text{ (at boundary)}} + \frac{1}{2} \mathbf{v}^\top H_z \mathbf{v}.
\end{equation} 
By $\mathbf{v}^\top H_z \mathbf{v} \geq \kappa ||\mathbf{v}||^2$, provided $||\mathbf{v}|| \leq \sqrt{\frac{2 |\mathcal{M}(z)|}{\kappa}}$  guarantee $\mathcal{M}(z + \mathbf{v}) \leq 0 \implies f(z + \mathbf{v}) \neq f(z)$ (the curvature condition following \cite{Wang2025-px}).  

Next, for $l\leq L/2$, the  discrete Fourier transform (DFT) satisfies
$|\mathrm{DFT}(\Theta(\hat{\delta}))_k| = 0 \quad \forall |k| > l/2.$ Thus, $\Theta(\hat{\delta})$ lies in the low-frequency subspace $\mathcal{F}_{\mathrm{low}} \subset \mathbb{R}^L$~\cite{Traversing}. By Jacobian alignment, $J_f(x) \cdot \Theta(\hat{\delta}) \in S$ prevails, i.e.,  
\begin{equation}
|| \mathrm{Proj}_S(J_f(x) \cdot \Theta(\hat{\delta})) || \geq c || J_f(x) \cdot \Theta(\hat{\delta}) ||.
\end{equation}  
We next define $\hat{\delta}$ as the solution to:
\begin{equation}
\label{eq:maxe}
\max_{\hat{\delta} \in \mathbb{R}^\ell, ||\hat{\delta}|| \leq \epsilon} \mathbb{E}_x\left[ \mathbf{v}^\top H_z \mathbf{v} \right], \quad \mathbf{v} = J_f(x) \cdot \Theta(\hat{\delta}).
\end{equation} 
Then, by the curvature condition and Hoeffding’s inequality\cite{hoeffding1994probability,AzumaMSFT}, for $m$ independent dimensions in $S$:  
\begin{equation}
\label{eq:proj}
\mathbb{P}_x\left(\mathbf{v}^\top H_z \mathbf{v} \geq \kappa ||\mathbf{v}||^2\right) \geq 1 - e^{\left(-\frac{m \kappa^2 \epsilon^2}{2 \sigma^2}\right)},
\end{equation} 
combining with Eq.\ref{eq:expansion}, the result follows:
\begin{equation}
\label{eq:bound}
\mathbb{P}_x(f(x + \Theta(\hat{\delta})) \neq f(x)) \geq 1 - \beta - e^{\left(-\frac{m \kappa^2 \epsilon^2}{2 \sigma^2}\right)}.
\end{equation}
\end{proof}

\subsection{Proposed method}
\label{s:method}
Our method consists of carefully designed loss functions, preprocessing techniques, and an optimized training procedure, which ensures the robustness and generalization of the UAP across various audio samples.
Following the logic in~\cite{AdvBias2020Auto} and the optimization of $\delta$ in Eq.~\ref{eq:maxe}  we therefore emphasize the UAP generation strategy based on $\mathcal{F}_{\mathrm{low}}$-restricted perturbations alongside the loss-function design.

\textbf{Cosine Similarity \myRed{Loss}.}
The first loss function minimizes the cosine similarity $\rho$ between the original audio $x$ and the perturbed audio $x + \delta$. This loss ensures that the adversarial perturbation effectively alters the speaker representation: 
\begin{equation}
L_{\text{fooling}} = \rho(f(x), f(x+\delta)).
\end{equation}

\textbf{The Exponential TV Loss.} We introduce a novel loss function inspired by the total variation (TV) loss commonly used in the image domain \cite{ZolfiAES22} to preserve the imperceptibility of the UAP. The loss function is designed as follows:
\begin{equation}
L_{\text{Exp TV}} = \frac{1}{l} \sum_{i=0}^{l-1} \phi(\hat{\delta_{i}}, \hat \delta_{i+1}),
\end{equation}
where $\delta_i$ UAP's amplitudes and
\begin{equation}
\phi(x, y) =
\begin{cases}
    \exp{(|y| - |x|)} - 1 & \text{if } |y| > |x|, \\
    \exp{(|y|)} - 1 & \text{elif } \text{sign}(y) \neq \text{sign}(x), \\
    0 & \text{else}.
\end{cases}
\end{equation}
Here, $l = 3200$ corresponds the UAP $\hat{\delta}$ length (0.2 seconds at a 16 kHz sampling rate). Our goal was to penalize only when the absolute values of the amplitudes increase, while avoiding penalization when they decrease. This setup allows the perturbation to remain minimal in less critical regions, while adapting to higher magnitudes where necessary. This pointwise approach distinguishes our loss function from traditional losses, enabling a more targeted and adaptive perturbation strategy. 

\textbf{Audio Preprocessing.}\label{s:preprocessing}
Due to the varying lengths of the audio samples in the dataset, we standardized them by padding each audio to a fixed length of 20 seconds. However, instead of using conventional zero-padding, which could lead to the UAP learning to attack the zero-padded portions, we opted to repeat the audio content to reach the target length. This approach ensures that the UAP focuses on the actual audio signal, preventing it from exploiting the artificially padded sections.

\begin{figure}[t]
  \centering
  \includegraphics[width=0.9\linewidth]{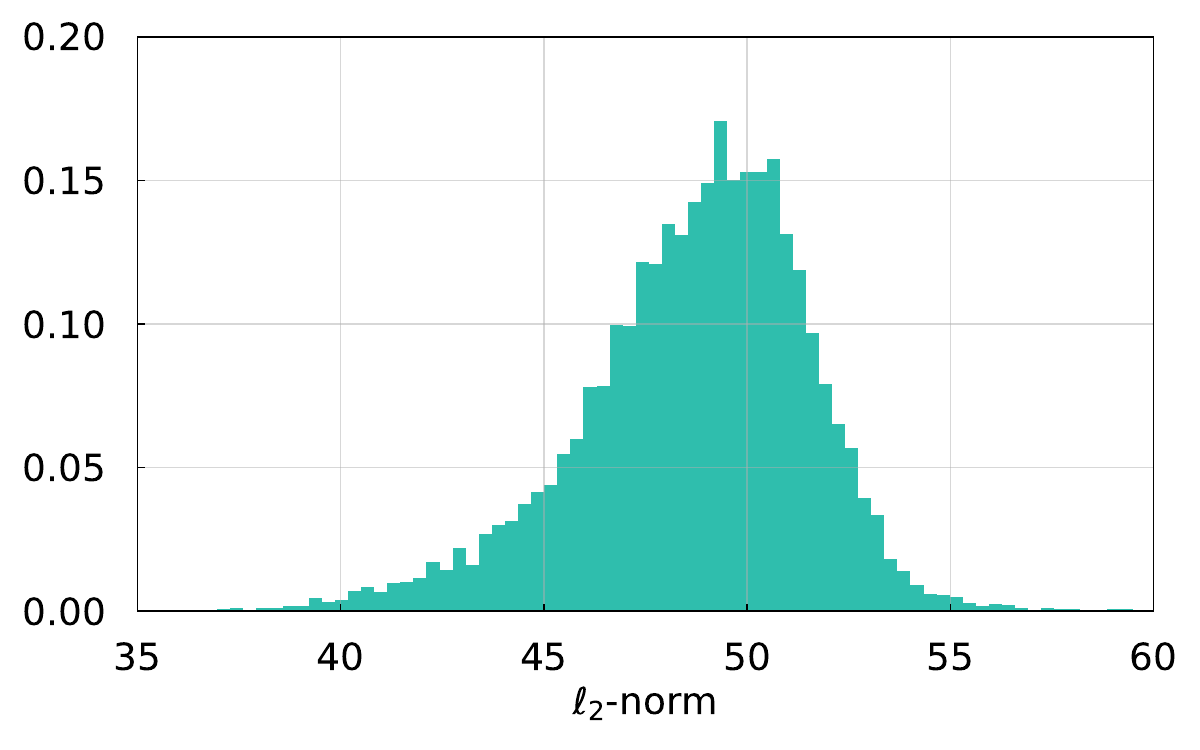}
  \caption{The distribution of loudness levels ($\ell_2$-norm) across the dataset.}
  \label{fig:cosine_similarity}
\end{figure}

Our analysis of the VoxCeleb2 dataset revealed variations in audio loudness, reflected in the $\ell_2$-norm distribution. These inconsistencies pose challenges for UAP generation, affecting perturbation perceptibility and test metrics. Figure~\ref{fig:cosine_similarity} highlights this issue, emphasizing the need for volume normalization to ensure uniform perturbation strength. To address this, we employed the \texttt{pyloudnorm} Python library \cite{steinmetz2021pyloudnorm} to standardize loudness levels across all audio samples.

\section{Experimental Setup}
We considered part of the VoxCeleb2 dataset \cite{chung2018voxceleb2} for our experiments. Each audio was normalized and padded to a 20-second duration using a repeat strategy, as detailed in Section \ref{s:preprocessing}. Our training set consists of approximately 61,000 audio samples and a validation set of 5,000 audio samples representing 500 distinct speakers \myRed{of diverse age, gender, language and accent}. Additionally, we used a test set of 10,000 audio samples from speakers not present in the training set. Unlike other works that often use shorter audio segments, we employed longer 20-second audio samples during training, allowing the UAP to be repeated multiple times within each sample. This approach helps the UAP adapt to anonymize audio of any length, improving the performance across various audio durations.

We employed the ECAPA-TDNN \cite{desplanques2020ecapa} model from the SpeechBrain framework \cite{ravanelli2021speechbrain} as the backbone for ASI and embedding extraction. This model is a strong baseline for the ASI and verification tasks. We chose the Whisper \myRed{v3 base} \cite{radford2023robust} model for ASR performance evaluation.

We considered three metrics to assess the effectiveness and imperceptibility of our UAP. \emph{Fooling Rate} is the percentage of audio samples for which the speaker recognition model produces incorrect predictions after applying the adversarial perturbation. The \emph{Signal-to-Noise Ratio (SNR)} metric quantifies the strength of the perturbation relative to the original audio signal.

Thirdly, \emph{Perceptual Evaluation of Speech Quality (PESQ)}, which evaluates the perceived quality of audio signals, is widely used in speech processing. It compares the original and perturbed audio, providing a score between -0.5 and 4.5, with higher values indicating better audio quality. Because it aligns closely with human perception, PESQ is considered more reliable than SNR when assessing speech quality. Finally, \emph{Word Error Rate, WER} evaluates the ASR performance on \myRed{processed} audio to estimate the effect of UAP on downstream applications. 

Our primary loss function is a weighted combination of the fooling loss ($w_1=1$) and the Exponential TV loss ($w_2 = 30$) that balances the fooling-imperceptibility trade-off. We employed the Adam \cite{kingma2014adam} optimizer with learning rate $3 \times 10^{-3}$ to generate the adversarial perturbation iteratively. We trained the UAP for $250$ epochs with batch size equal to $64$. To prevent excessive UAP magnitudes, we applied a clipping operation with a predefined threshold $\epsilon = 0.01$ after each update. Given that the length of \myRed{processed} audio cannot be predetermined, we will generate a UAP with a fixed duration of 0.2 seconds. This perturbation will then be repeated continuously until it matches the length of the target audio.

\section{Results and Discussion}

In this section, we present the results of our experiments, analyse their impact, and compare our approach with existing method. 

\textbf{Exponential TV Loss Evaluations.} To assess the impact of our proposed Exponential TV loss, we compare two different UAPs: one trained using our loss and another trained with the conventional $\ell_2$-norm loss, commonly used by competitors, e.g., \cite{israel} for the SAS training. The $\ell_2$-norm loss is defined as $L_{\ell_2} = \frac{||\hat{\delta}||_2}{l}$, where $l$ is the UAP length. 

The results of our comparison are presented in Table \ref{tab:boltzmann_vs_l2}. For a \myRed{proper} evaluation, we set the maximum length of \myRed{the} test audio samples \myRed{at} $20$ seconds. However, in that case, shorter audio samples were not padded, ensuring that the UAP is assessed under natural conditions without introducing artificial padding effects. This approach prevents any advantage or disadvantage that could arise from padding shorter audio samples.

\begin{table}[th]
  \caption{Comparison of our methods with the state-of-the-art.}
  \label{tab:boltzmann_vs_l2}
  \centering
  \begin{tabular}{l r r r r}
    \toprule
    Loss Function & FR (\%) & SNR & PESQ & WER (\%) \\
    \midrule
    Ours, $L_{\ell_2}$         & 74.80 & 17.54 & 2.48 & 94.1 \\
    Ours, $L_{\text{Exp TV}}$ & 75.70 & 19.18 & 2.68 & 52.6 \\
    S.Hanina et al.\cite{israel} & 67.94 & 19.37 & 2.74 & 73.2 \\
    \bottomrule
  \end{tabular}
\end{table}

Our results demonstrate that the Exponential TV loss effectively enhances the trade-off between fooling rate and imperceptibility. While achieving a fooling rate comparable to the conventional $\ell_2$-norm loss, our method improves imperceptibility by yielding a higher SNR and a better PESQ score. This demonstrates that UAP generated using the $L_{\text{Exp TV}}$ are more effective and perceptually less noticeable than those produced with the traditional $\ell_2$-loss. The results support the probabilistic guarantee for adversarial patch success, with FR bounded below by Eq.~\ref{eq:bound}. The use of a $\mathcal{F}_{\mathrm{low}}$-subspace ensures more imperceptible perturbations as reflected in high SNR and aligned with Eq.\ref{eq:proj}, restricting perturbations to perceptually better regions.

\begin{figure}[t]
  \centering
  \includegraphics[width=0.8\linewidth]{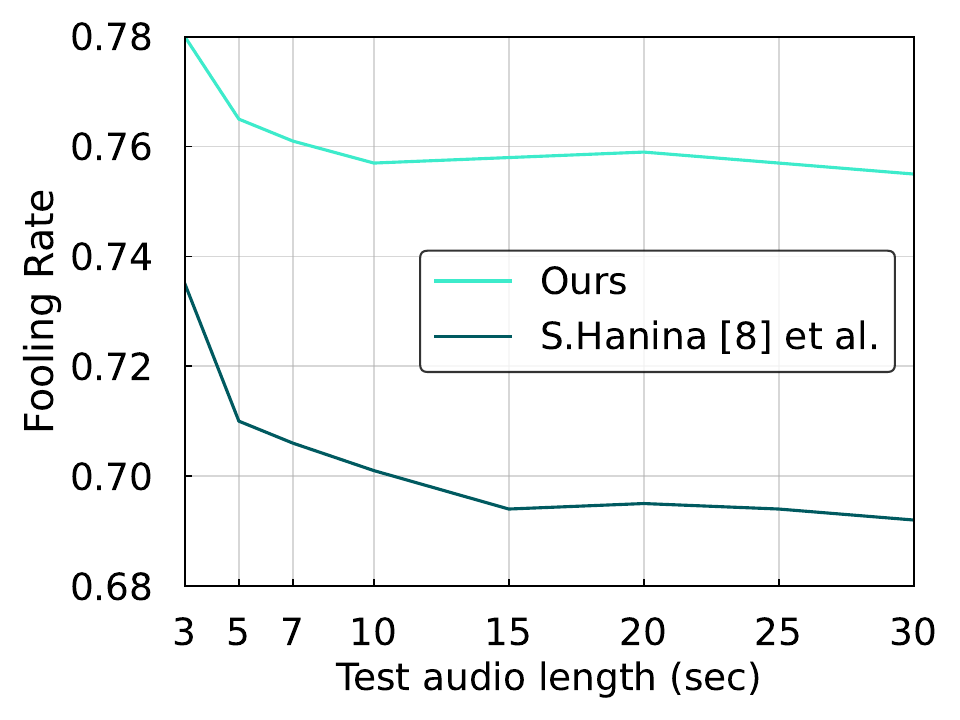}
  \caption{Comparison of the proposed UAP (with $L_{\text{Exp TV}}$) approach performance with that of~\cite{israel} across different test audio lengths.}
  \label{fig:len_test}
\end{figure}

\textbf{Evaluating the dependence of UAP on the audio length.}
Previous works \cite{israel, waveunet} have focused on training and testing their UAPs on fixed-length audio samples, leading to overfitting to a specific duration. This contradicts the fundamental definition of a universal perturbation, which should be generalised across varying audio lengths. This limitation is partly due to the training strategy used in prior approaches, where the UAP length is set equal to that of the training audio. As a result, the UAP does not repeat over longer audio segments and fails to adapt effectively to different durations.  
\begin{figure}[t]
  \centering
  \includegraphics[width=0.85\linewidth]{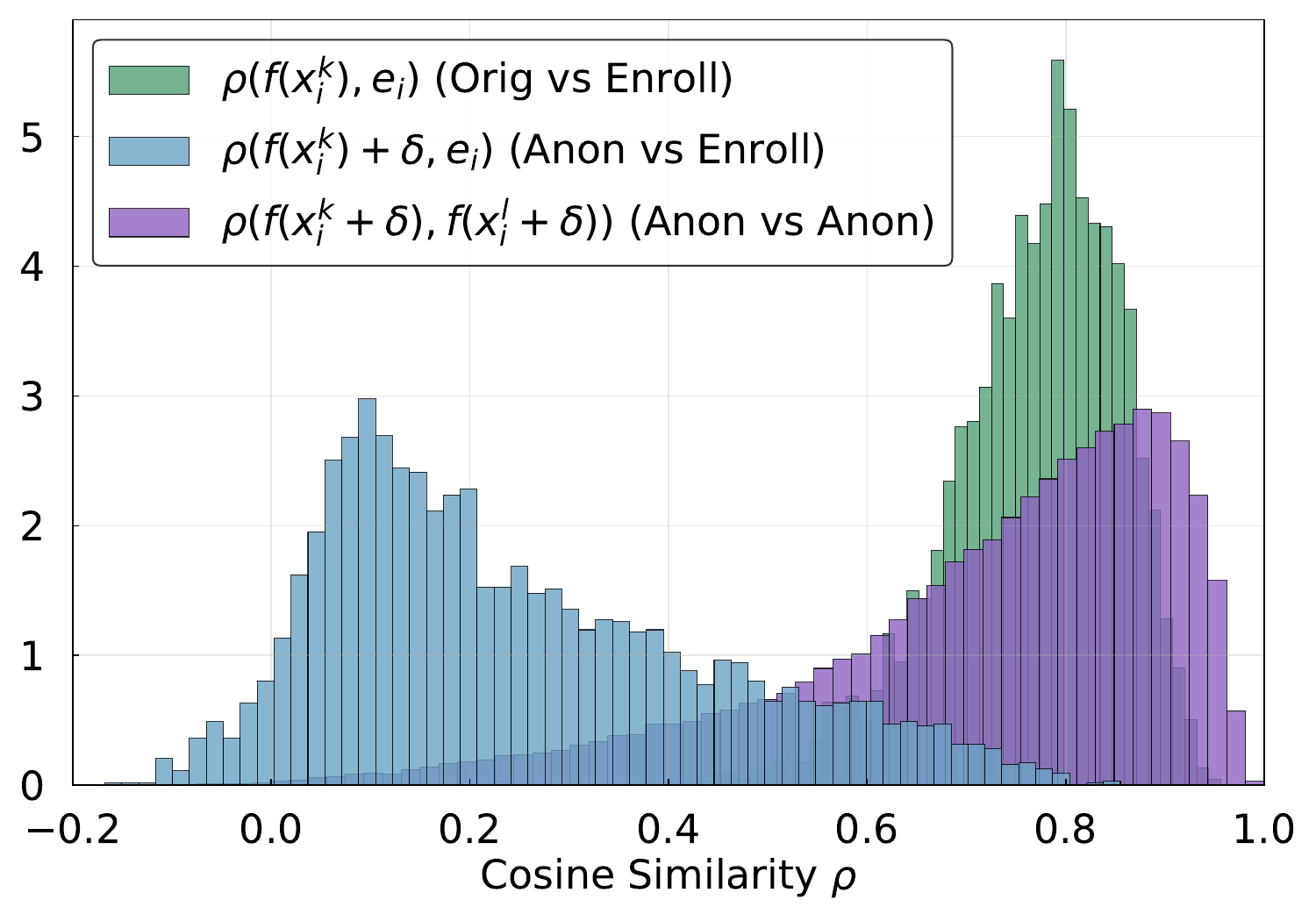}
  \caption{Cosine similarity distributions between original (''Orig'') audio and Enrolment vectors (''Enroll''), between anonymized audio (''Anon'') and Enrolment vectors, and between different anonymized audio of the same speakers. }
  \label{fig:embeddings_sim}
\end{figure}
Moreover, the methods in \cite{liu2024transferable, LiZJXZWM020} that utilize short UAPs and repeat them across the audio length still train and test on audio of the same fixed duration, further restricting generalization. We evaluate UAP performance across different test audio lengths to address this issue, ensuring a more comprehensive and robust assessment.

Figure \ref{fig:len_test} shows that our UAP outperforms the UAP from \cite{israel} across all audio lengths. Notably, our UAP maintains higher robustness on longer audio samples, with only a $2\%$ drop in fooling rate, compared to more than $4\%$ for the alternative approach. Another important insight from these results is that misleading the SR model is easier when testing on short audio samples, such as 3 seconds. This suggests that testing on short audio samples does not fully assess the capabilities of the SR model.

\textbf{Embeddings Similarity Analysis.} We analysed the distribution of cosine similarities to assess the impact of the UAP on speaker embeddings, as shown in Figure \ref{fig:embeddings_sim}. We specifically computed the cosine similarity between three pairs of vectors: (1) embeddings of original non-attacked audio samples and corresponding enrolment vector of the speaker, (2) adversarially perturbed and original audio samples from the same speaker, and (3) two different adversarially perturbed samples from the same speaker. The results were averaged across all speakers. We used five audio to create an enrollment vector and 20 audio to evaluate all 75 test speakers to collect the presented statistics.

The speaker recognition model initially performs well, with high cosine similarity for same-speaker audio. UAP application reduces similarity between original and perturbed samples, showing attack effectiveness, though some samples remain resistant. High similarity between different perturbed samples of the same speaker indicates the ASI model maps anonymized audio to closely clustered vectors per speaker.

\section{Conclusion and Future Work}

In this study, we proposed a novel approach to voice anonymization by developing and evaluating UAPs with a new loss function designed to enhance the trade-off between imperceptibility and fooling rate. Our method demonstrated superior performance in fooling rate, perceptual quality, and robustness on long-duration audio compared to existing approaches. We additionally introduced a \myRed{length-agnostic} evaluation to ensure a reliable assessment of adversarial attacks. Future work might focus on improving UAP generalization to various voice conditions, biometrics models, and real-world application scenarios.

\newpage

\bibliographystyle{IEEEtran}
\bibliography{mybib}

\begin{thebibliography}{10}
\providecommand{\url}[1]{#1}
\csname url@samestyle\endcsname
\providecommand{\newblock}{\relax}
\providecommand{\bibinfo}[2]{#2}
\providecommand{\BIBentrySTDinterwordspacing}{\spaceskip=0pt\relax}
\providecommand{\BIBentryALTinterwordstretchfactor}{4}
\providecommand{\BIBentryALTinterwordspacing}{\spaceskip=\fontdimen2\font plus
\BIBentryALTinterwordstretchfactor\fontdimen3\font minus \fontdimen4\font\relax}
\providecommand{\BIBforeignlanguage}[2]{{%
\expandafter\ifx\csname l@#1\endcsname\relax
\typeout{** WARNING: IEEEtran.bst: No hyphenation pattern has been}%
\typeout{** loaded for the language `#1'. Using the pattern for}%
\typeout{** the default language instead.}%
\else
\language=\csname l@#1\endcsname
\fi
#2}}
\providecommand{\BIBdecl}{\relax}
\BIBdecl

\bibitem{desplanques2020ecapa}
B.~Desplanques, J.~Thienpondt, and K.~Demuynck, ``{ECAPA-TDNN}: Emphasized channel attention, propagation and aggregation in {TDNN} based speaker verification,'' in \emph{INTERSPEECH}, 2020, pp. 3830--3834.

\bibitem{wang2023cam++}
H.~Wang, S.~Zheng, Y.~Chen, L.~Cheng, and Q.~Chen, ``Cam++: A fast and efficient network for speaker verification using context-aware masking,'' in \emph{Proc. INTERSPEECH}, 2023, pp. 5301--5305.

\bibitem{deng2023v}
J.~Deng \emph{et~al.}, ``V-cloak: Intelligibility-, naturalness- \& timbre-preserving real-time voice anonymization.'' in \emph{Proc. USENIX Security Symposium}.\hskip 1em plus 0.5em minus 0.4em\relax USENIX Association, 2023, pp. 5181--5198.

\bibitem{xie2021enabling}
Y.~Xie \emph{et~al.}, ``Enabling fast and universal audio adversarial attack using generative model,'' in \emph{Proceedings of the AAAI conference on artificial intelligence}, 2021.

\bibitem{zhang2023imperceptible}
X.~Zhang, X.~Zhang, M.~Sun, X.~Zou, K.~Chen, and N.~Yu, ``Imperceptible black-box waveform-level adversarial attack towards automatic speaker recognition,'' \emph{Complex \& Intelligent Systems}, vol.~9, no.~1, pp. 65--79, 2023.

\bibitem{szegedy2013intriguing}
C.~Szegedy \emph{et~al.}, ``Intriguing properties of neural networks,'' in \emph{Proc. ICLR}, 2014.

\bibitem{Korzh2025}
D.~Korzh \emph{et~al.}, ``Certification of speaker recognition models to additive perturbations,'' \emph{Proc. AAAI}, vol.~39, no.~17, pp. 17\,947--17\,956, 2025.

\bibitem{moosavi2017universal}
Moosavi-Dezfooli \emph{et~al.}, ``Universal adversarial perturbations,'' in \emph{Proc. CVPR}, 2017, pp. 1765--1773.

\bibitem{israel}
S.~Hanina, A.~Zolfi, Y.~Elovici, and A.~Shabtai, ``Universal adversarial attack against speaker recognition models,'' in \emph{Proc. ICASSP}.\hskip 1em plus 0.5em minus 0.4em\relax IEEE, 2024, pp. 4860--4864.

\bibitem{snyder2018x}
D.~Snyder \emph{et~al.}, ``X-vectors: Robust dnn embeddings for speaker recognition,'' in \emph{Proc. ICASSP}.\hskip 1em plus 0.5em minus 0.4em\relax IEEE, 2018, pp. 5329--5333.

\bibitem{jung2019RawNet}
J.-w. Jung \emph{et~al.}, ``Rawnet: Advanced end-to-end deep neural network using raw waveforms for text-independent speaker verification,'' \emph{Proc. INTERSPEECH}, pp. 1268--1272, 2019.

\bibitem{Borodin2024}
K.~Borodin \emph{et~al.}, ``Aasist3: Kan-enhanced aasist speech deepfake detection using ssl features and additional regularization for the asvspoof 2024 challenge,'' in \emph{Proc. ASVspoof}.\hskip 1em plus 0.5em minus 0.4em\relax ISCA, 2024, p. 48–55.

\bibitem{wang2023wespeaker}
H.~Wang \emph{et~al.}, ``Wespeaker: A research and production oriented speaker embedding learning toolkit,'' in \emph{Proc. ICASSP}.\hskip 1em plus 0.5em minus 0.4em\relax IEEE, 2023, pp. 1--5.

\bibitem{liu2024transferable}
X.~Liu, H.~Tan, J.~Zhang, A.~Li, and Z.~Gu, ``Transferable universal adversarial perturbations against speaker recognition systems,'' \emph{World Wide Web}, vol.~27, no.~3, p.~33, 2024.

\bibitem{NeekharaHPDMK19}
P.~Neekhara \emph{et~al.}, ``Universal adversarial perturbations for speech recognition systems,'' in \emph{INTERSPEECH}, 2019, pp. 481--485.

\bibitem{ZhangDTWG23}
J.~Zhang \emph{et~al.}, ``Daob: A transferable adversarial attack via boundary information for speaker recognition systems.'' in \emph{IJCAI}, 2023, pp. 12--17.

\bibitem{LuoSLX21}
H.~Luo \emph{et~al.}, ``Spoofing speaker verification system by adversarial examples leveraging the generalized speaker difference,'' \emph{Sec. and Commun. Netw.}, vol. 2021, Jan. 2021.

\bibitem{waveunet}
Y.~Xie \emph{et~al.}, ``Enabling fast and universal audio adversarial attack using generative model,'' in \emph{Proc. AAAI}, 2021.

\bibitem{gulati2020conformer}
A.~Gulati \emph{et~al.}, ``Conformer: Convolution-augmented transformer for speech recognition,'' in \emph{Proc. INTERSPEECH}, 2020, pp. 5036--5040.

\bibitem{radford2023robust}
A.~Radford, J.~W. Kim, T.~Xu, G.~Brockman, C.~McLeavey, and I.~Sutskever, ``Robust speech recognition via large-scale weak supervision,'' in \emph{ICML 2023}.\hskip 1em plus 0.5em minus 0.4em\relax PMLR, 2023, pp. 28\,492--28\,518.

\bibitem{shen2018natural}
J.~Shen \emph{et~al.}, ``Natural tts synthesis by conditioning wavenet on mel spectrogram predictions,'' in \emph{ICASSP 2018}.\hskip 1em plus 0.5em minus 0.4em\relax IEEE, 2018, pp. 4779--4783.

\bibitem{miao2023speaker}
X.~Miao, X.~Wang, E.~Cooper, J.~Yamagishi, and N.~Tomashenko, ``Speaker anonymization using orthogonal householder neural network,'' \emph{IEEE/ACM Transactions on Audio, Speech, and Language Processing}, 2023.

\bibitem{miao2022analyzing}
X.~Miao \emph{et~al.}, ``Analyzing language-independent speaker anonymization framework under unseen conditions,'' in \emph{INTERSPEECH}, 2022, pp. 4426--4430.

\bibitem{wang2023vsmask}
Y.~Wang, H.~Guo, G.~Wang, B.~Chen, and Q.~Yan, ``Vsmask: Defending against voice synthesis attack via real-time predictive perturbation,'' in \emph{Proc. ACM WiSec}, 2023, pp. 239--250.

\bibitem{goodfellow2014explaining}
I.~J. Goodfellow, J.~Shlens, and C.~Szegedy, ``Explaining and harnessing adversarial examples,'' in \emph{Proc. ICLR}, Y.~Bengio and Y.~LeCun, Eds., 2015.

\bibitem{madry2017towards}
A.~Madry, A.~Makelov, L.~Schmidt, D.~Tsipras, and A.~Vladu, ``Towards deep learning models resistant to adversarial attacks,'' in \emph{Proc. ICLR}, 2018.

\bibitem{Rahmati2020GeoDAAG}
A.~Rahmati, S.-M. Moosavi-Dezfooli, P.~Frossard, and H.~Dai, ``Geoda: a geometric framework for black-box adversarial attacks,'' in \emph{Proc. CVPR}, 2020, pp. 8446--8455.

\bibitem{Wang2025-px}
B.~Wang, ``\BIBforeignlanguage{en}{Starshaped compact hypersurfaces in warped product manifolds {II}: A class of hessian type equations},'' \emph{\BIBforeignlanguage{en}{J. Geom. Anal.}}, vol.~35, no.~1, Jan. 2025.

\bibitem{Traversing}
J.~Bayer, S.~Becker, D.~M\"{u}nch, M.~Arens, and J.~Beyerer, ``Traversing the subspace of adversarial patches,'' \emph{Mach. Vision Appl.}, vol.~36, no.~3, 2025.

\bibitem{hoeffding1994probability}
W.~Hoeffding, ``Probability inequalities for sums of bounded random variables,'' in \emph{The collected works of Wassily Hoeffding}.\hskip 1em plus 0.5em minus 0.4em\relax Springer, 1994, pp. 409--426.

\bibitem{AzumaMSFT}
O.~Dekel and O.~Shamir, ``There's a hole in my data space: Piecewise predictors for heterogeneous learning problems,'' in \emph{Proc. AISTATS}, vol.~22.\hskip 1em plus 0.5em minus 0.4em\relax PMLR, 21--23 Apr 2012, pp. 291--298.

\bibitem{AdvBias2020Auto}
A.~Liu \emph{et~al.}, ``Bias-based universal adversarial patch attack for automatic check-out,'' in \emph{ECCV}.\hskip 1em plus 0.5em minus 0.4em\relax Berlin, Heidelberg: Springer-Verlag, 2020, p. 395–410.

\bibitem{ZolfiAES22}
A.~Zolfi, S.~Avidan, Y.~Elovici, and A.~Shabtai, ``Adversarial mask: Real-world universal adversarial attack on face recognition models,'' in \emph{Proc. ECML PKDD}.\hskip 1em plus 0.5em minus 0.4em\relax Springer, 2022, pp. 304--320.

\bibitem{steinmetz2021pyloudnorm}
C.~J. Steinmetz and J.~D. Reiss, ``pyloudnorm: {A} simple yet flexible loudness meter in python,'' in \emph{150th AES Convention}, 2021.

\bibitem{chung2018voxceleb2}
J.~S. Chung, A.~Nagrani, and A.~Zisserman, ``Voxceleb2: Deep speaker recognition,'' in \emph{Proc. INTERSPEECH}, 2018, pp. 1086--1090.

\bibitem{ravanelli2021speechbrain}
M.~Ravanelli \emph{et~al.}, ``Speechbrain: A general-purpose speech toolkit,'' \emph{arXiv preprint arXiv:2106.04624}, 2021.

\bibitem{kingma2014adam}
D.~P. Kingma and J.~Ba, ``Adam: A method for stochastic optimization,'' in \emph{Proc. ICLR}, 2015.

\bibitem{LiZJXZWM020}
J.~Li, X.~Zhang, C.~Jia, J.~Xu, L.~Zhang, Y.~Wang, S.~Ma, and W.~Gao, ``Universal adversarial perturbations generative network for speaker recognition,'' in \emph{Proc. IEEE ICME}, 2020.

\end{thebibliography}
\end{document}